\documentclass[preprint2]{aastex}

\shorttitle{Testing a pulsating binary model for long secondary
periods in red variables}
\shortauthors{Jundan Nie et al.}
\begin{document}
\title{Testing a pulsating binary model for long secondary
periods in red variables\\}
\author{J.D.~Nie}
\affil{Department of Astronomy, Beijing Normal University,
     Beijing 100875, China}
\email{niejundan@mail.bnu.edu.cn}

\author{X.B.~Zhang}
\affil{National Astronomical Observatories, Chinese Academy of
Sciences, Beijing 100012, China} \email{xzhang@bao.ac.cn}

\author{B.W.~Jiang}
\affil{Department of Astronomy, Beijing Normal University,
     Beijing 100875, China}
\email{bjiang@bnu.edu.cn}

\begin{abstract}
The origin of the long secondary periods (LSPs) in red variables
remains a mystery up to now, although there exist many models. The
light curves of some LSPs stars mimic an eclipsing binary with a
pulsating red giant component. To test this hypothesis, the
observational data of two LSP variable red giants, 77.7795.29 and
77.8031.42, discovered by the MACHO project from the LMC, are
collected and analyzed. The probable eclipsing features of the light
curves are simulated by the Wilson-Devinney (W-D) method. The
simulation yields a contact and a semidetached geometry for the two
systems, respectively. In addition, the pulsation constant of the
main pulsating component in each binary system is derived. By
combining the results of the binary model and the pulsation
component, we investigate the feasibility of the pulsating binary
model. It is found that the radial velocity curve expected from the
binary model has a much larger amplitude than the observed one and a
period double the observed one. Furthermore, the masses of the
components based on the density derived from the binary orbit
solution are too low to be compatible with both the evolutionary
stage and the high luminosity. Although the pulsation mode
identified by the pulsation constant which is dependent on the
density from the binary-model is consistent with the first or second
overtone radial pulsation, we conclude that the pulsating binary
model is a defective model for the LSP.
\end{abstract}


\section{INTRODUCTION}
Among the variable red giant stars, one sub-type exhibits long
secondary periods (LSPs). The light curves of these stars exhibit
not only a short primary period but also  a long secondary period,
which is approximately nine times longer than the short one. This
phenomenon has been known for several decades \citep{pay54,hou63}.
Some samples of these LSPs variables are shown in \citet{kis99}. An
interest in the stars with the LSPs has been renewed by the study of
\citet{woo99}. This paper shows that in the LMC, $\sim$25~\% of all
variable asymptotic giant branch (AGB) stars show LSP roughly nine
times longer than the short primary period which is typically $\sim$
30--200 days. Meanwhile, a study of the bright local pulsating red
giants indicates that at least one-third of these stars exhibit LSPs
\citep{per04}. Soszynski also gives $\sim$30\% of pulsating red
giants in the LMC with LSPs \citep{sos07b}. In the period-luminosity
(P-L) diagram, it is interesting to see that the LSP variables
follow a distinct sequence (sequence D), which is roughly parallel
to the radial pulsation sequences A, B, and C for variable red
giants.

The LSPs present in variable red giants  have attracted a lot of
attention since their discovery, but their origin still remains
mysterious. Since the LSPs are several times  longer than the
fundamental radial periods, they could not be caused by normal
radial pulsations. Moreover, \citet{woo99} and \citet{woo04} note
that the LSPs can not be explained by $g^{+}$ mode for the
oscillatory $g ^{+}$ mode is evanescent in convective region and it
is unlikely to be observable in a red giant because of its
convective envelope. The $g^{-}$ mode is dynamically unstable in the
convection envelope, and unable to lead to any oscillation.
Regarding the nonradial $p$ modes, their periods are rather shorter
than those of the fundamental radial modes, so they can not explain
the LSPs either. Alternatively, if the LSPs are caused by some
strange modes, they would be extremely damped and should not be seen
\citep{woo00}. One more possible explanation is the rotating
spheroid model, which can explain the shape of the velocity curve,
but there is no reason for the rotating period to bring about the
observed P-L relation.

It seems that variable red giants with long secondary periods
(hereinafter referred to as LSPVs) can not be easily interpreted as
pulsating red giants. In this situation, a hypothesis of binarity
arose. It has been suggested that the sequence D stars could be
components of close binary systems, and the LSPs could be
interpreted as the light variations caused by ellipsoidal binary
motions or eclipses. \citet{sos04,sos07b} find that the sequence D
stars overlap with and have a direct continuation of the sequence E
stars that are mostly confirmed to be binaries. Radial velocity
variations on the time-scale of LSPs have also been measured in a
number of sequence D stars, such radial velocity variations being a
requirement if the sequence D stars are binaries
\citep{hin02,oli03,woo04,nic09}. Some observational arguments
supporting the binary hypothesis have already been reviewed by
\citet{sos07a}.

After examining the light curves of LSPVs collected from the MACHO
database, we note that many of them show some nearly regular and
stable, eclipse-like light variations at the long secondary periods
with a large amplitude in comparison with that of the primary
pulsation. It seems that the LSPs can be easily interpreted as
eclipses by an orbiting component. If this can be shown to be the
case, it would be direct evidence to support the binary hypothesis
for the sequence D stars. To test this idea, we propose a model --
pulsating binary, i.e. a binary system with a pulsating component.
We begin with some very probable eclipsing LSPs candidates whose
light curves look like eclipsing binary with primary and secondary
minima. The light curves of two LSP stars are collected and analyzed
by using the Wilson-Devinney ( W-D, hereafter) code and the
power-spectrum method. Afterwards, the proposed eclipsing and
pulsating nature as well as their evolutionary properties are
discussed.

\section{THE DATA}

Since the LSPs are sometimes as long as 1500 days, the candidate
selection is based on the long-term MACHO survey. The MACHO
project is a microlensing experiment that monitored numerous stars
in the LMC, SMC and Galactic Bulge over a time of $\sim$ 3000 days.
As the by-products of this survey, a large number of variables were
discovered and monitored, including many LSPVs. From this
database, two LSPVs in the LMC are selected as our program objects.
Their F.T.S (field.title.sequence) numbers are 77.7795.29 and
77.8031.42, respectively. They are chosen as the working sample for
the following reasons: (1) The time base of observations for both of
the stars is long enough to cover at least  a couple of LSPs. (2)
Their light variations are both eclipse-like and symmetric over the
long secondary period, and have distinct short primary variation.
(3) These two stars are quite different from each other not only in
their LSPs but also in the shapes of the light curves, they are
expected to be different types of binaries.

The MACHO data are taken in non-standard red and blue bandpasses.
The adopted transformation from the MACHO instrumental photometry to
Kron-Cousins $R$ and $V$ are given in \citet{alc99}. However, it
does not work well for our candidates in practice and brings about
visible dispersion for the photometric points. Analyzing the data
sets, we find the reason mainly being that the observed color index
term ``$(V_{M,t}-R_{M,t})$'' in \citet{alc99} brings in large
dispersion as the raw red and blue magnitudes transfer their errors
to each other. This term requires both measurements to be good at
each time when adopting the transformation. In addition, the
transformation has a term involving airmass. When the value of the
airmass is greater than 2.0, the correction for the atmosphere
extinction is rather uncertain. We adopt another transformation to
avoid the above problems as \citet{str06}:
\begin{mathletters}
\begin{eqnarray}
\ R=R_{M,t} +23.90+0.1825(V_{M,t}-R_{M,t})  \\
 V=V_{M,t} +24.22-0.1804(V_{M,t}-R_{M,t})
  ,
\end{eqnarray}
\end{mathletters}
where the term ``$V_{M,t}-R_{M,t}$'' is  the mean color index for a
red giant instead of an individual measurement. Since color indexes
of the red giants have a small scattering of about 0.2 magnitude,
the adoption of the mean multiplied by a factor of 0.18 brings about
a dispersion of only 0.04 mag, much smaller than the photometric
uncertainty. The light curves of standardized $R$ and $V$ magnitudes
for our two candidates are shown in Fig.~1.

OGLE--II \citep{uda97,szy05} also provides the light variation data
for red giants in the $I$ band, and the time span of the light curve
overlaps the last half of the MACHO data, with $\sim$ 300 days
extension which is not long  for the LSP. Besides and more
importantly, the data are much sparser than the MACHO in the
overlapping time range and it is of little help in analyzing the
short primary pulsation. Thus we don't include the OGLE--II data for
following analysis.

\section{THE BINARY MODEL}

Assuming that the phenomenon of the LSPs is caused by an
orbiting component, we can employ some binary simulation method
to analyze the LSPs light curve. The W-D method, which consists
of two main FORTRAN programs -- LC and DC (short for Light
Curve and Differential Correction, respectively), is a
simulation method producing the photometric solution for the
binary system.

The 2007 version of the W-D code is used to analyze the
eclipse-like light curves in $R$ and $V$ bandpasses of our two
candidates. Nonlinear limb-darkening via a logarithmic form
along with many other features \citep{wil71,wil79,wil90,kal98}
are used in the code. Considering the likely close distance
between the components, the effect of reflection is taken into
account.

In computing the photometric solution, the important parameters
adopted in the DC program are as follows. The orbital period for
adjustment is computed by the PDM (Phase Dispersion Minimization)
method \citep{ste78} and the Period04 software for double-check at
the beginning. The value is about 970 days for 77.7795.29 and 1700
days for 77.8031.42. Note that these orbital periods are twice the
normally adopted value for the LSP, the latter corresponding to one
cycle of long period variation apparent in the light curve. The
temperatures of the two primary stars are both set to 3311 K
according to \citet{fox82} for long period stars. The initial
bolometric (X1,X2,Y1,Y2) and monochromatic (x1,y1,x2,y2)
limb-darkening coefficients of the components are taken from
\citet{van93}. The gravity darkening exponents are set to 0.32 for
both the primary and the secondary component for convective
envelopes according to \citet{luc67}. The bolometric albedos are
taken as A1=0.4 and A2=0.5 following \citet{ruc69}.

The preliminarily estimated parameters are put in the first
iteration of the DC code and after several iterations, a converged
solution is reached. The photometric solutions are shown in Table 1,
and the synthesized $R$ band light curves computed from the LC code
are shown in Fig.~1 together with the observed curves of the two
candidates. The photometric solutions in the table provide insight
into the orbital dynamics of the binary system. The orbit period
($P_{\rm orb}$) adjusted by the DC code is consistent with the value
from the PDM and Period04 method. The large angles of inclination
($i$), and argument of periastron ($\omega$) describe the
orientation of the binary orbit, while the value of $e$ reveals a
nearly circular orbit. Moreover, we find that the secondary
component is a low mass M-type star, according to the mass ratio and
effective temperature. All the binary modes, including detached
mode, semidetached mode, and contact mode in the code are tried in
the solution-seeking procedure, and it is found that only one mode
can reach a converged solution for each candidate. For 77.7795.29, a
contact configuration converges, and for 77.8031.42, a semidetached
system with the primary component filling its Roche lobe converges.
This implies that both of them have a mass exchange with their
components due to Roche lobe overflow. No converged solutions were
found for systems showing purely ellipsoidal variations i.e. systems
with an invisible secondary companion.

Using the photometric solutions, the theoretical radial velocities
can be also simulated by the LC code. We try this in order to
compare them with some observational facts. The bottom panel of
Fig.~1 shows the synthesized velocity curves for the primary
components. The full amplitude of the velocity variation is about
11.0 km~s$^{-1}$  for 77.7795.29,  and 9.0 km~s$^{-1}$ for
77.8031.42, where the value of the binary system mass is taken from
later Section 5 (but see discussion there for the problem of mass).
Yet all the studies of radial velocity for LSPVs
\citep{hin02,oli03,woo04,nic09} show typical amplitudes of 3-4
km~s$^{-1}$ and no objects with full velocity amplitude greater than
7 km~s$^{-1}$ are found. Moreover, the synthesized velocity curves
of the primary components have a period which is twice of the light
variation period. In contrast, all the observed radial velocities
for the LSPVs have the same period as that of the light variation.
The synthesized radial velocities for the second stars are also
obtained from the LC code, and we find that the full amplitude is
about 18.5 km~s$^{-1}$ for 77.7795.29 and 18.3 km~s$^{-1}$ for
77.8031.42. The large velocity separations between the primary and
secondary stars would have been readily seen, while in the existing
spectral observations there is no indication of this. Therefore the
synthesized velocity curve based on the binary model is inconsistent
with the observation.

In addition, we should note that, among all the parameters derived
from the DC code, some are highly reliable, including the orbital
period, mass ratio and temperature difference if we admit the binary
hypothesis, and some are relatively uncertain, such as the effective
temperatures of the two components. Several parameters derived from
the LC code, such as the mass and radius, are particularly uncertain
for there are no radial velocities available for comparison, so they
are not listed in Table~1. The synthesized radial velocities
produced by the LC code are also very uncertain.

\section{THE INTRINSIC PULSATION}

For an eclipsing binary with a pulsating (primary) component, the
observed light curve could be approximately interpreted as
\citep{zha09}:
\begin{equation}
 l_{\rm obs}= l_{1}f_{\rm pul}+l_{2},
\end{equation}
where $l_{1}$ and $l_{2}$ are the calculated brightness contributed
by the primary and secondary components respectively, and $f_{\rm
pul}$ denotes the pulsating variation of $l_{1}$. With the derived
binary photometric solution, we could compute the brightness of the
binary components separately at each observation epoch by using the
LC code in the W-D code package, from which the ``pure'' pulsation
light variation of the primary star could then be extracted.
However, we do not use this approach, because the intrinsic
pulsating amplitude of the primary component is so large that the
theoretically synthesized light curve computed by the LC code would
have up-down fluctuation and the extracted light curve would not be
exactly equivalent to the true intrinsic pulsation variations.
Instead, we make use of the original data sets to analyze the
primary pulsation.

Period04, a commonly used technique for time series analysis, which
utilizes Fourier transforms as well as multiple-least-squares
algorithms, is applied to the power spectrum analysis. In doing
that, we use the $R$ band time series data which are brighter and
with higher photometric accuracy than in the $V$ band. We select
only those peaks in the power spectrum with signal to noise ratio
(S/N) larger than 4.0 for further discussion. Fig.~2 and Fig.~3 show
the first four-frequency solution of the data sets in the $R$ band
before and after subtraction of the most prominent remaining
frequency (prewhitening). At the top of the two figures the spectral
window based on the epoches of available observations is displayed.
We notice that the alias patterns, including the 1 c/d daily alias,
are quite low in power. The next four patterns in both figures are
the step-by-step power spectrum derived from the pulsation data. The
fitting is stopped when the light curve is well fitted. For
77.7795.29, nine frequencies are obtained, by using most parts of
the MACHO data while the last part of the MACHO data is dropped
because of a large gap. Among all the nine frequencies obtained, we
find that the main power spectrum are dominated at $f_1=0.0020$ c/d
and $f_2=0.0100$ c/d. $f_1$ is two times the assumed orbital
frequency ($f_0=1/970=0.00103$ c/d), it is not accepted as the real
pulsating frequency, so only $f_2$ remains as the intrinsic
pulsating frequency. The other derived frequencies are found to be
related to $f_0$ and $f_2$ as follows:  $f_1=2f_0$, $f_3=f_2+6f_0$,
$f_4=f_0$, $f_5=4f_0$, $f_6=f_2+1/365$, $f_7=f_2+7f_0$, $f_8=3f_0$,
$f_9=5f_0$, all of them are aliases. For the other star 77.8031.42,
we obtain $f_2=0.0112$ c/d, the real pulsating frequency, and the
others have the relations of: $f_1=2f_0$, $f_3=4f_0$,
$f_4=f_2+0.0004$, $f_5=3f_6$, $f_6=2f_0+f_8$, $f_7=4f_6$,
$f_8=1/365$, $f_9=f_2+10f_0$, $f_{10}=f_2-0.0004$, $f_{11}=f_6+f_8$,
$f_{12}=10f_0$, $f_{13}=12f_0$, $f_{14}=f_0$, where $f_0$ is the
assumed orbital frequency. Here, 0.0004 c/d, may be the rotation
frequency of the LSPVs, in the theoretical range of 2400--10000 days
for the rotation period of red giants \citep{kis00,woo04}. Based on
the above analysis, only one short primary pulsation is found for
our two candidates respectively and the result is consistent with
previous result from \citet{woo99} for LSPs stars.

The eclipsing light curve synthesis has provided us some important
parameters of the binary system such as the mass ratio and the
unified radius $r_1=R_1/A$ which normalizes the radius to the
semi-major axis (actually the accurate ratio can be obtained by the
LC code). With these values, the mean density of the pulsating
primary component can be precisely determined. From the Kepler's law
\begin{equation}
 P_{\rm orb}^{2}= \frac{4\pi^{2}A^{3}}{G(M_1+ M_2)}
\end{equation}
and the density in solar units
\begin{equation}
\rho_{1}/\rho_{\sun} = \frac{M_{1}/R_{1}^3}{M_{\sun}/R_{\sun}^3} ,
\end{equation}
we get:
\begin{equation}
\rho_{1}/\rho_{\sun}=
\frac{4\pi^{2}R_{\sun}^{3}}{M_{\sun}G(1+q)r_{1}^{3}P_{\rm orb}^2},
\end{equation}
where $P_{\rm orb}$ is the orbit period of the binary system.

Substituting the values of $P_{\rm orb}$, $q$ and $r_1$ from
the W-D code analysis of the light curve into Eq.~(5), the mean
density is computed, being $8.47\times10^{-8}$ and
$3.67\times10^{-8}$ in solar unit respectively, which agrees
with the density of red giants.  Then the pulsation constant
can be easily calculated from:
$Q=P_{\rm{pul}}(\rho_{1}/\rho_{\sun})^{1/2}$, where $P_{\rm
pul}$ is the ``pure'' pulsation period. The pulsation constant
$Q$ is an important parameter to tell the intrinsic oscillation
mode of a pulsating star, especially in normal radial mode. The
$Q$ value calculated on the binary hypothesis, is shown in
Table~3 and equal to 0.029 and 0.017 for our two candidates,
respectively.

The values of $r1$ calculated by the LC code in Table 3, 0.47
and 0.44 respectively, are consistent with that estimated in
\citet{sos07b} which notices $R_{1}/A\approx0.4$ in the binary
scenario. The result of mode identification shows agreement
with the theoretical expectation for red giants \citep{fox82}
and reveals that 77.7795.29 is a first overtone pulsating star
while 77.8031.42 is a second overtone pulsating star. The
pulsation properties derived from a binary model are reasonable
for a pulsating red giant.

\section{THE EVOLUTIONARY PROPERTY}
A further analysis of the pulsating binary model is carried out
from the evolutionary property of the components, as the
stellar properties of the red giants are rarely known, by
making use of the results derived in Section 4. From the 2MASS
near-infrared photometric data and the empirical bolometric
correction factor $BC_{\rm K}$ as a function of $(J-K)$
\citep{bes84}, and with a distance modulus of 18.54 for the
LMC, the bolometric magnitude can be calculated, and the
luminosity can thus be obtained. Applying the formula $L=4\pi
\sigma R^{2}T_{\rm {eff}}^{4}$, the equilibrium radius of the
primary star can be computed. Here, $T_{\rm{eff}}$ is an
assumed value, 3311 K, which is the mean temperature for long
period variables in LMC. We do not calculate the temperature
from the color index because the secondary component is not
understood well and its contribution to the observed color
index is not known. With the mean density and the radius, the
mass of the primary star is derived. All these values are
listed in Table 3.

The luminosity of the primary stars in the table is a few thousands
of solar luminosity, consistent with the red giant luminosity. The
radii of both candidate stars are on the order of hundreds of solar
radius, which are consistent with red giants. However, the masses of
the two LSPVs are too low, and the masses of their components would
be even lower, according to the mass ratios in Table 1. Moreover,
these low mass stars have very high luminosities, which seems
impossible. For the star 77.7795.29, the mass is 0.31$M_{\sun}$ and
its luminosity is 2502$L_{\sun}$. In the age of universe, such a low
mass star is impossible to evolve up to the red giant branch. For
the secondary component, the mass is about 0.188$M_{\sun}$, while
its luminosity is 1607$L_{\sun}$, the value of which is computed
from the data of ``$L_{1}/(L_{1}+L_{2})$'' in Table~1. It is hard to
imagine how such low mass stars in a binary system exchange their
masses. It is difficult to find an evolutionary path to satisfy this
situation. This problem is also present in the star 77.8031.42, with
a mass of 0.36$M_{\sun}$  and a luminosity of 4839$L_{\sun}$, whose
component has about 0.175$M_{\sun}$ and 3790$L_{\sun}$. If a
reasonable mass is expected, the density should be upgraded by a
factor of four or the radius be upgraded by a factor of 1.4.
Recalling the determination of the density from Eq.~(5), the density
is calculated from the parameters, $q$, $r_1$ and $P_{\rm orb}$ all
of which have relatively small errors which will not bring about a
density uncertainty of a factor of four. Besides, the density is
reasonable for a red giant. On the other hand, the radius is derived
from the luminosity and effective temperature. The luminosity has
uncertainty from the bolometric correction and the neglect of
interstellar extinction, and the effective temperature also has some
uncertainty. If the luminosity is higher or the effective
temperature lower, the radius would be larger and so the mass.
However, if the mass is indeed higher, the system would then have a
large separation from Eq.(3) and an even more serious problem with
the large velocity amplitude. In summary, the pulsating binary model
requires a low mass incompatible with an evolved stellar stage
having the observed high luminosity.

\section{DISCUSSION AND CONCLUSION}
The origin of the LSPs is unknown and there exist many explanations.
The light curves of some LSPVs are eclipse-like and it seems that
this phenomenon is due to an invisible component orbiting around the
pulsating red giant. To test this hypothesis, we propose a model --
pulsating binary, and select two LSP stars to analyze their orbit
motions and pulsation nature.

On the assumption of binarity, we simulate the photometric light
curves of the systems by using the W-D method. The photometric
solutions give us a configuration for the binary system: a contact
system for 77.7795.29 and a semidetached system for 77.8031.42 with
its Roche lobe fully filled. It means that the LSP star may have
strong interaction and very probably mass transfer with the other
component via Roche lobe overflow. However, Roche lobe overflow will
rule out the binary hypothesis if we accept the view of
\citet{woo04}, which argues that the mass transfer of Roche lobe
will result in a short merger timescale of about 1000 yrs. Moreover,
the calculated effective temperatures and the
``$L_{1}/(L_{1}+L_{2})$'' values in Table 1 suggest that the
secondary star is also a red giant with a similar temperature and
luminosity to that of the primary red giant. This would lead to a
double-lined spectroscopic binary. However, no observer of radial
velocities in these systems has reported seeing spectral lines from
the secondary star \citep{hin02,oli03,woo04,nic09}. There are also
some serious problems about the synthesized velocity curves. The
simulation produces one cycle of velocity curve for two cycles of
the light curve, while the observed velocities show one cycle of the
radial velocity curve for one cycle of the light curve. The
synthesized full velocity amplitudes for the star 77.7797.29 and
77.8031.42 are much greater than the typical values of other LSPVs.
The full amplitudes of the secondary components are even larger and
would lead to  great velocity separations between the primary and
the secondary stars, which have never been found by spectral
observations.

Fourier analysis is applied to investigate the intrinsic oscillation
of the LSPVs over the raw data sets. Using the parameters ($R_{1}/A$
and $q$) obtained from the W-D code, the mean densities of the LSPVs
are deduced and they are consistent with the red giant phase. Then
the pulsation constants are obtained by using the classical equation
$Q=P_{\rm{pul}}(\rho_{1}/\rho_{\sun})^{1/2}$ to identify the
pulsation mode. For star 77.7795.29, one pulsating frequency is
detected and it is caused by the first overtone radial pulsation,
and for star 77.8031.42, the only pulsating frequency is caused by
the second overtone radial pulsation. These agree with the
theoretical value for red giants and  the conclusion of
\citet{woo99}.

The stellar properties of LSPVs are derived by using the information
from the pulsating binary model. We calculate the bolometric
magnitude, luminosity, radius and mass of the primary star, and find
some of them conflict very seriously with the evolutionary
properties of red giants. In particular, the masses for the star
77.7795.29 and 77.8031.42 are both less than 0.4$M_{\sun}$, and
their luminosities are too high for such low masses. It is difficult
to imagine how such  low mass stars could have such large
luminosities. The situation gets even worse if we apply the mass
ratio and the value of ``$L_{1}/(L_{1}+L_{2})$'' to calculate the
mass and luminosity of the secondary star.

Therefore, the radial velocities  and the masses computed from the
pulsating binary model do not agree with some observations and facts
about red giants. We conclude that the model ``pulsating binary''
has some deficiencies in dealing with the observed properties of
LSPVs and that the binary hypothesis for explaining the LSPs seems
unreasonable.

\acknowledgments The authors are very grateful to Prof. Peter
Wood for very constructive suggestions and helpful discussion.
They also thank the anonymous referee for helpful suggestions
to improve the work significantly. This research is supported
by the National Natural Science Foundation of China (NSFC)
through grant 10778601 and the Ministry of Science and
Technology of the People's Republic of China through grant
2007CB815406.

\clearpage

\begin{figure}
\begin{center}
\includegraphics[angle=90,width=0.85\hsize,height=0.60\hsize]{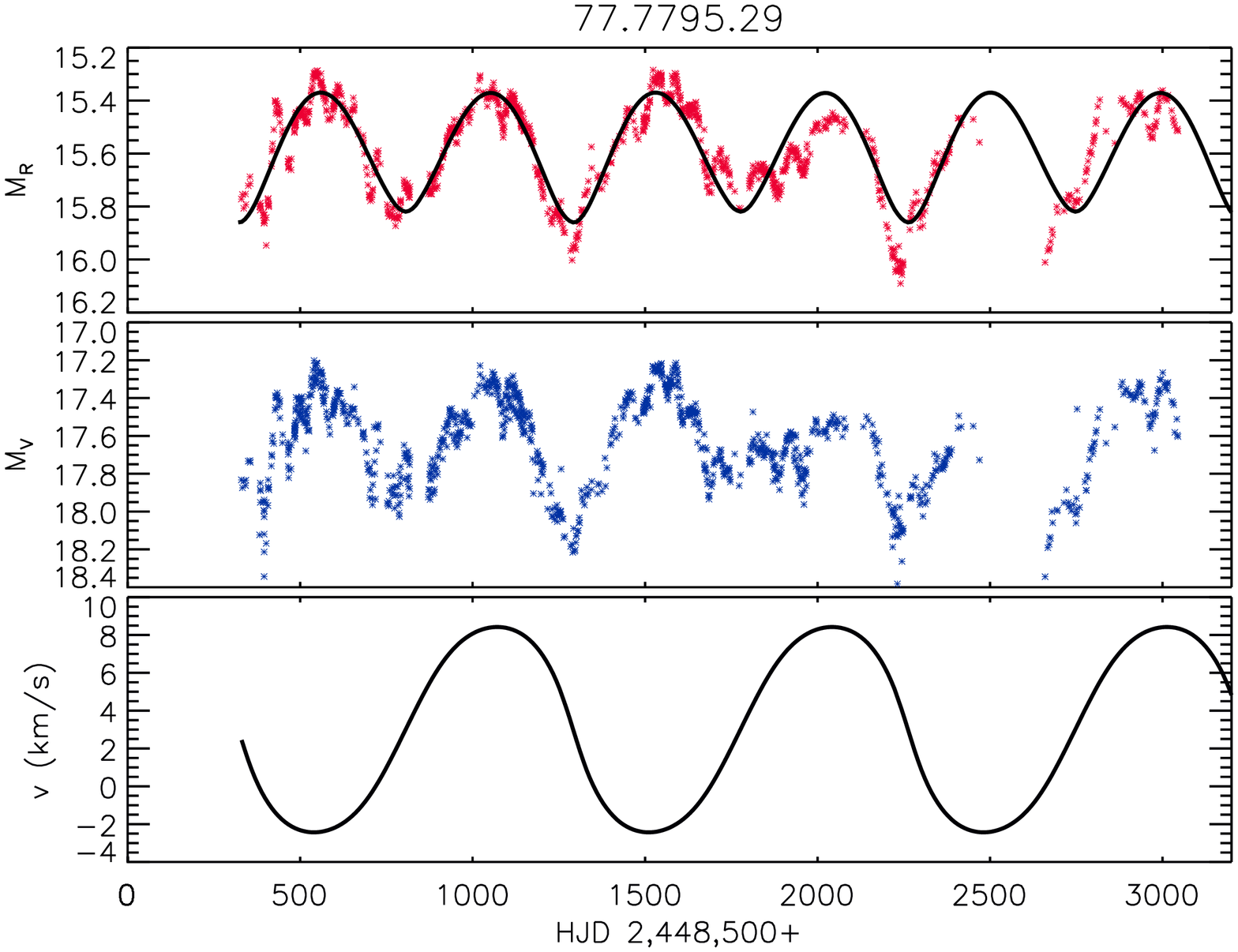}
\\[14pt]
\includegraphics[angle=90,width=0.85\hsize,height=0.60\hsize]{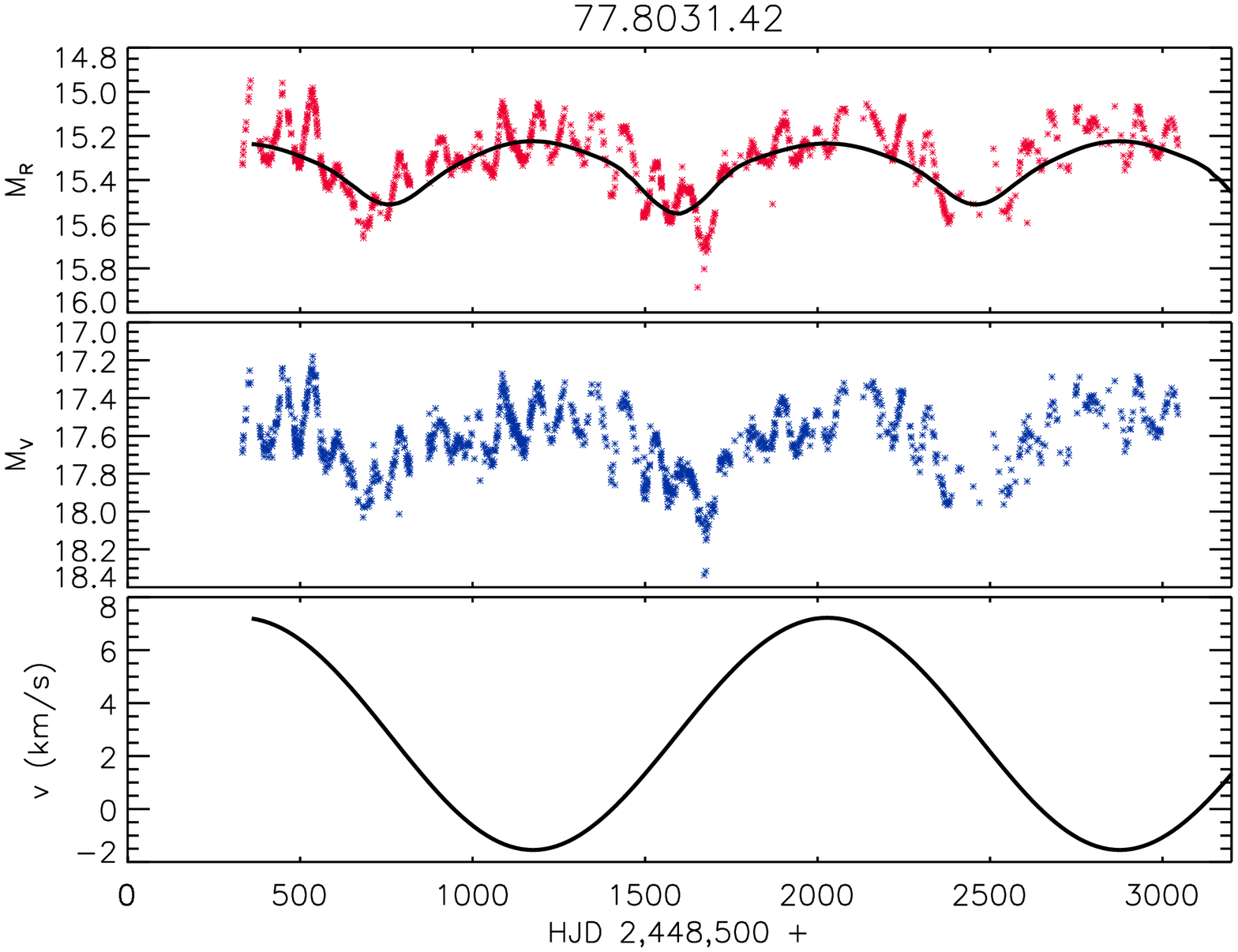}

\caption{Light and velocity curves synthesized by the W-D code
for 77.7795.29 and 77.8031.42. Also shown are the observed $R$
and $V$ band light curves from the MACHO project. Red asterisk
represents the observed data point in $R$ band while the blue
one is the $V$ band data point. The black solid line in the $R$
band panel denotes the theoretical synthesized light curve. The
bottom panel is the synthesized velocity curve.}
\end{center}
\end{figure}

\begin{figure}
\begin{center}
\includegraphics[angle=90,width=0.70\hsize,height=0.18\hsize]{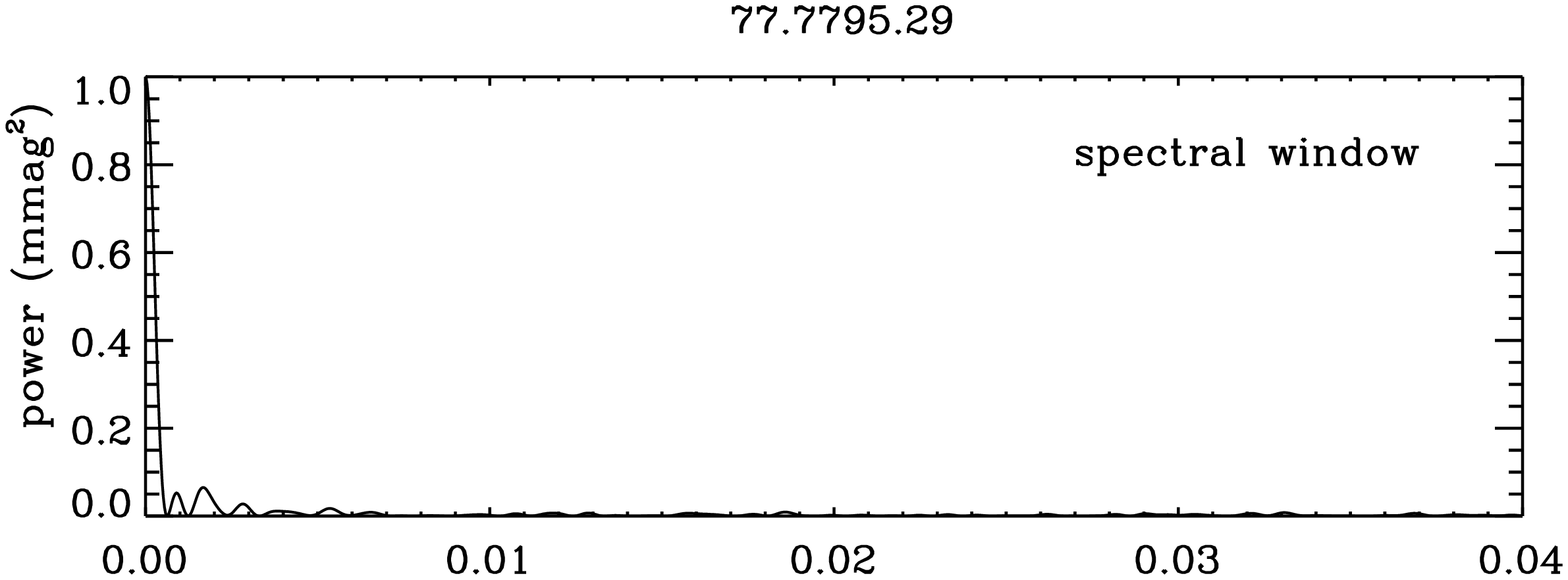}
\includegraphics[angle=90,width=0.7\hsize,height=0.55\hsize]{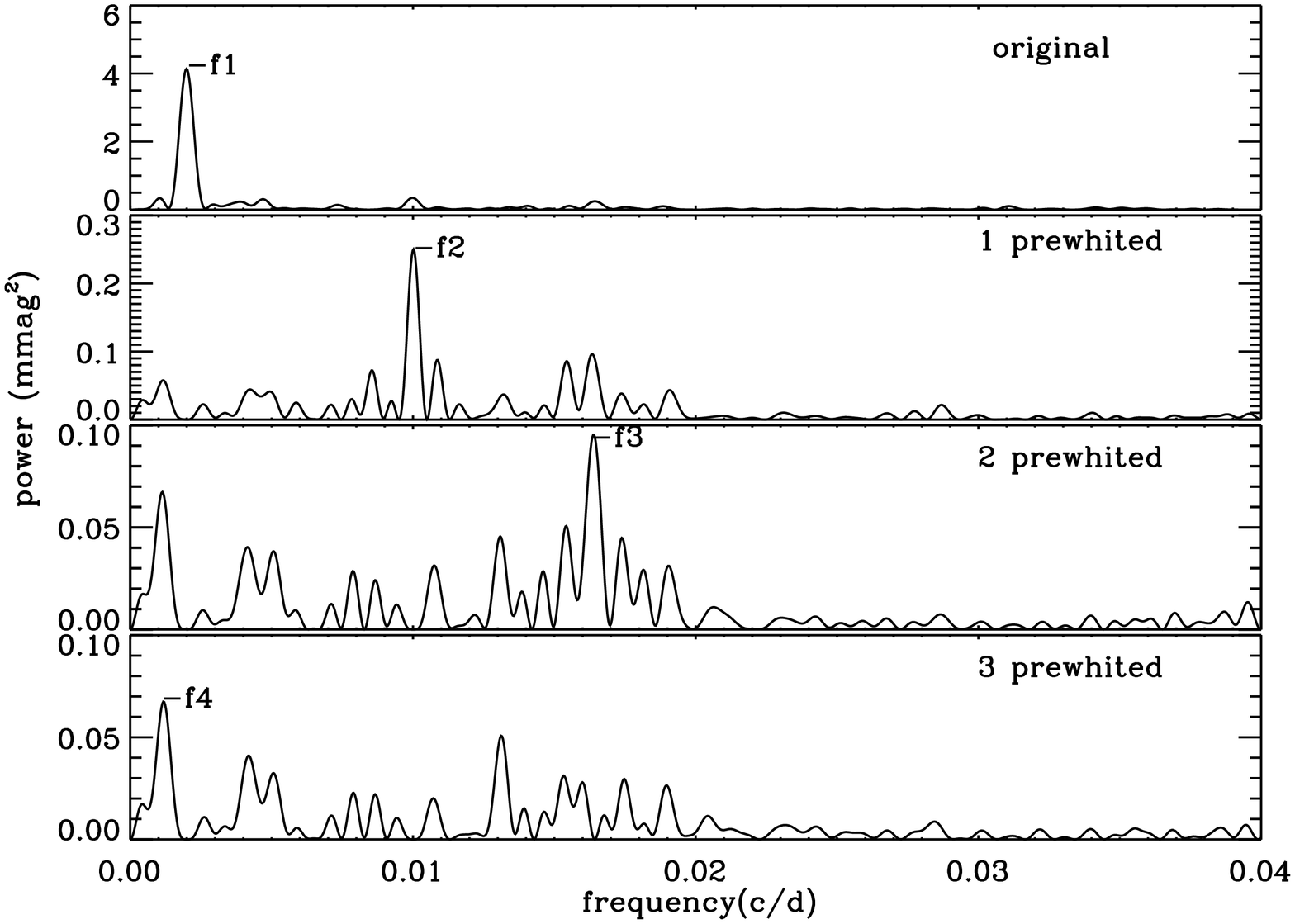}
\\[14pt]
\includegraphics[angle=90,width=0.71\hsize,height=0.25\hsize]{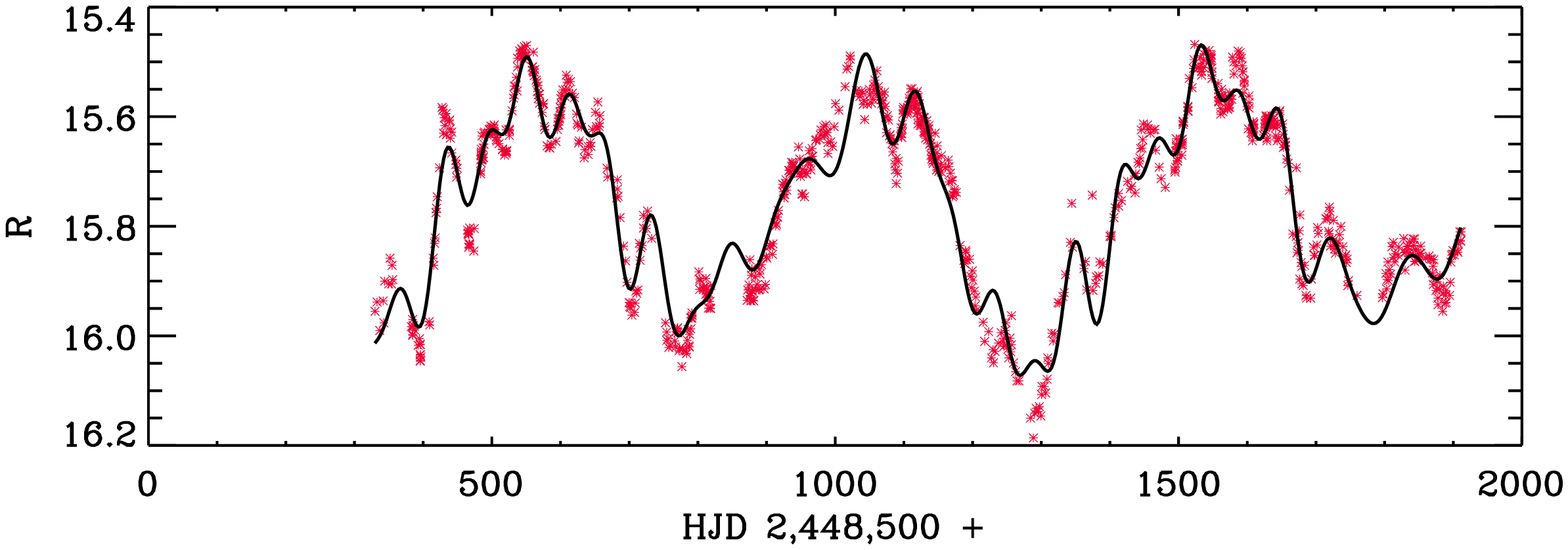}

\caption{Power spectrum and light curve of the primary star
77.7795.29 in binary system. The top five panels show power
spectra and the bottom panel shows  the light curve in the $R$
band. The black solid light curve is the fit of frequency
solution described in Sect.~4 of the text. }
\end{center}
\end{figure}

\begin{figure}
\begin{center}
\includegraphics[angle=90,width=0.70\hsize,height=0.18\hsize]{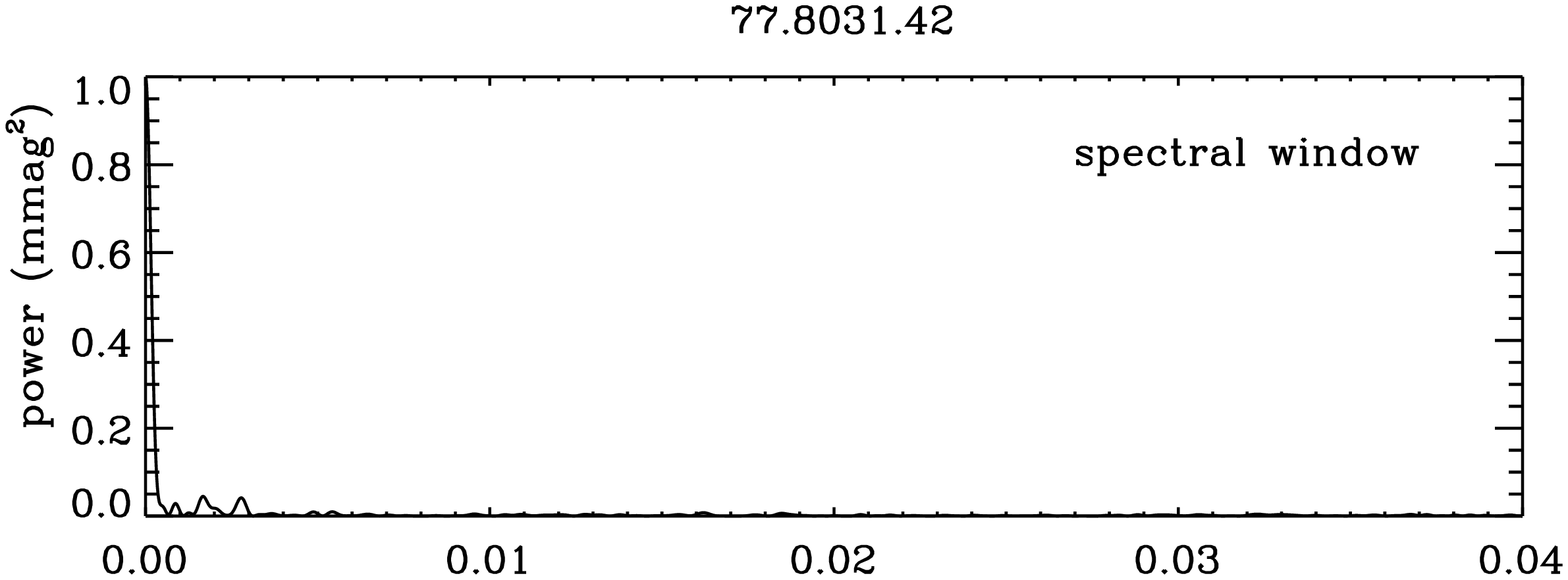}
\includegraphics[angle=90,width=0.7\hsize,height=0.55\hsize]{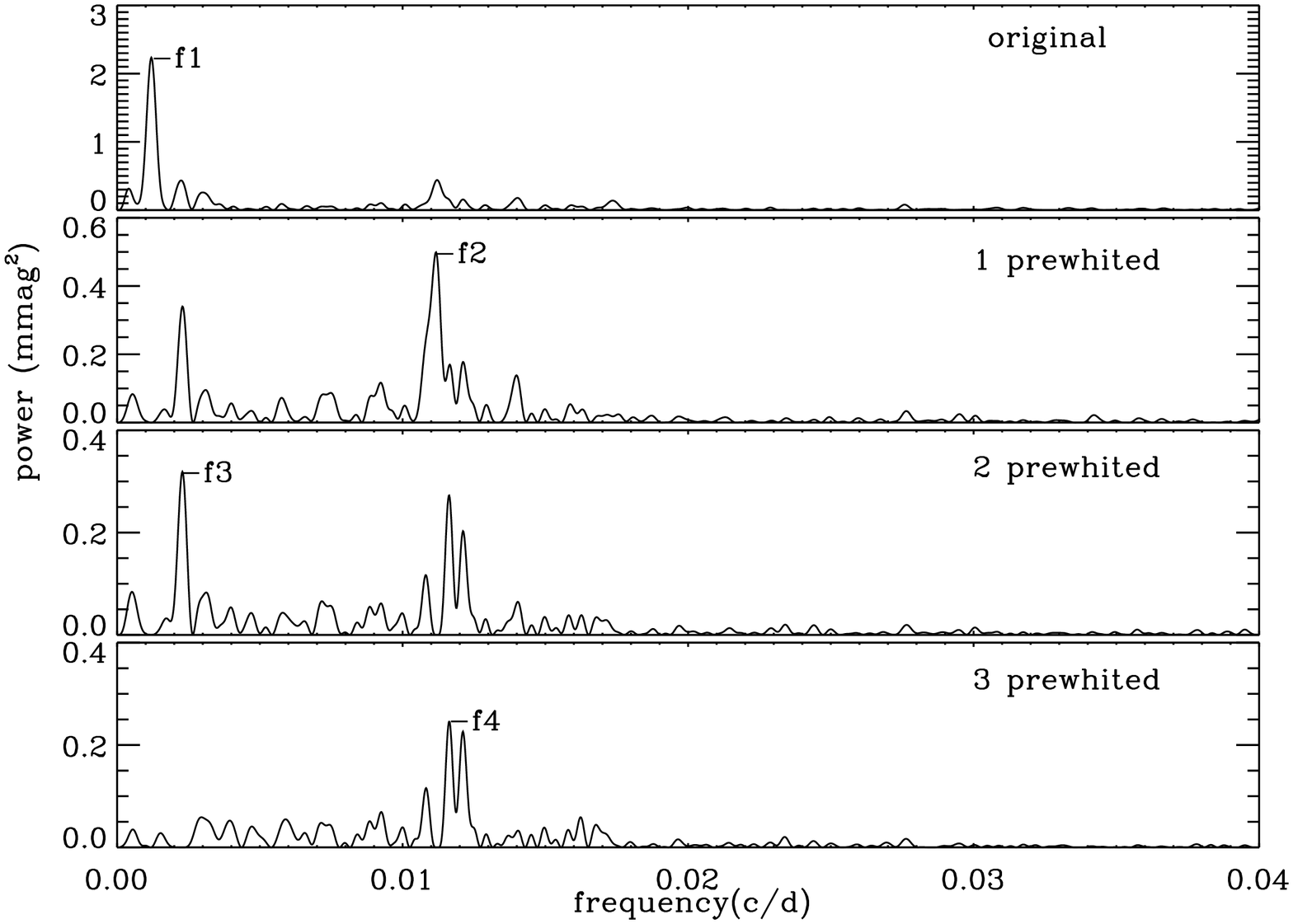}
\\[14pt]
\includegraphics[angle=90,width=0.7\hsize,height=0.25\hsize]{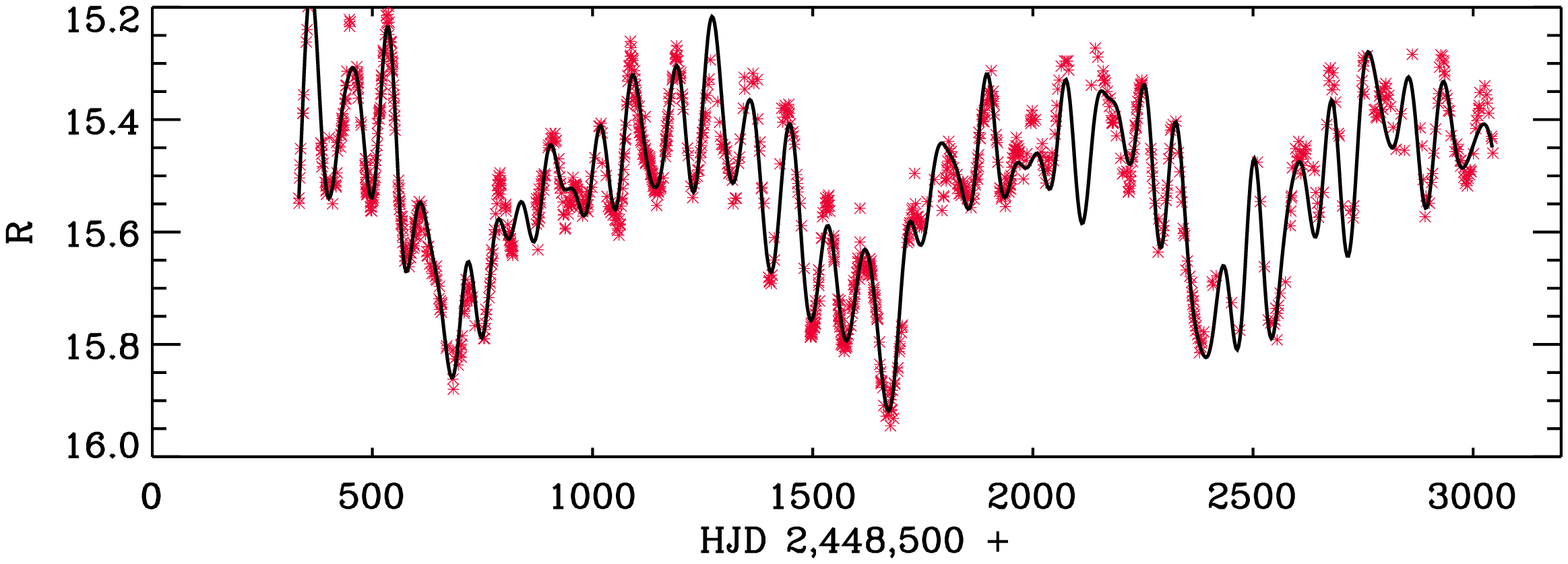}

\caption{The same as Fig.~2, but for the star 77.8031.42. }
\end{center}
\end{figure}
\clearpage

\begin{deluxetable}{clll}
\tabletypesize{\scriptsize} \tablecaption{Photometric solution for
stars 77.7795.29 and 77.8031.42.\label{tbl-1}} \tablewidth{0pt}
\tablehead{ \colhead{ } & \colhead{ \small{77.7795.29} } &
\colhead{} & \colhead{\small{77.8031.42}} } \startdata
\small{classification}
& \small{contact} && \small{semidetached}\\
\small$P_{\rm orb}$ (day) &\small970.9 &&\small1700.9\\
\small$T_{1}$ (K) &\small3311(assumed) &&\small3311(assumed) \\
\small$T_{2}$ (K) &\small3284  &&\small3544$\pm{48}$  \\
\small$e$ &\small0.0016$\pm{0.0008}$ &&\small0.0123$\pm{0.0164}$ \\
\small$i$ (degree) &\small62.49$\pm{0.27}$ &&\small63.19$\pm{2.22}$ \\
\small$\omega$ &\small3.64$\pm{0.20}$ &&\small2.91 \\
\small$q=m_{2}/m_{1}$ &\small0.608$\pm{0.003}$ &&\small0.486$\pm{0.029}$ \\
\small$\varphi_{1}$ &\small2.8359$\pm{0.0041}$ &&\small2.8748 \\
\small$\varphi_{2}$ &\small2.8359 &&\small2.8943 $\pm{0.055}$\\
\small$L_{1}/(L_{1}+L_{2})$  (V band) &\small0.609 &&\small0.561 \\
\enddata
\end{deluxetable}
\clearpage

\begin{deluxetable}{ccccc|cccc}
\tabletypesize{\scriptsize} \tablecaption{Results of Fourier
analysis for stars 77.7795.29 and 77.8031.42.\label{tbl-2}}
\tablewidth{0pt}

\tablehead{ \colhead{ } & \colhead{}& \colhead{\small{77.7795.29} }
& \colhead{} & \colhead{ } & \colhead{}& \colhead{\small{77.8031.42}
} & \colhead{} & \colhead{}} \startdata \small{F} &
\small{Frequency} & \small{Amplitude}&\small{Phase} & \small{S/N} &
\small{Frequency}
&\small{Amplitude} & \small{Phase} & \small{S/N}\\
& \small{(c/d)}&\small{(mag)} & &  & \small{(c/d)} & \small{(mag)} &
&\\
\hline
\small$f_{1}$& \small{0.0020}& \small0.2138& \small0.3144& \small21.6729& \small0.0011& \small0.1464&\small0.46905& \small19.3482 \\
\small$f_2$& \small0.0100& \small0.0529& \small0.6576& \small22.4815& \small0.0112& \small0.0770&\small0.7190& \small44.7148\\
\small$f_3$& \small0.0162& \small0.0327& \small0.0015& \small13.9081& \small0.0022& \small0.0613&\small0.2773& \small35.6114\\
\small$f_4$& \small0.0011& \small0.0363& \small0.9538& \small15.4348& \small0.0116& \small0.0446&\small0.7607& \small25.9384\\
\small$f_5$& \small0.0041& \small0.0280& \small0.6792& \small11.9004& \small0.0120& \small0.0404&\small0.6926& \small23.4930\\
\small$f_6$& \small0.0130& \small0.0205& \small0.8848& \small8.7069& \small0.0040& \small0.0260&\small0.1102& \small15.1249\\
\small$f_7$& \small0.0174& \small0.0189& \small0.4320& \small8.1379& \small0.0162& \small0.0267&\small0.1793& \small15.4983\\
\small$f_8$& \small0.0031& \small0.0199& \small0.0318& \small8.0322& \small0.0029& \small0.0351&\small0.8120& \small20.3622\\
\small$f_9$& \small0.0049& \small0.0199& \small0.6315& \small8.4640& \small0.0169& \small0.0224&\small0.8329& \small13.0511\\
\small$f_{10}$&\nodata &\nodata &\nodata &\nodata & \small0.0108& \small0.0311& \small0.7704&\small18.1114\\
\small$f_{11}$&\nodata& \nodata& \nodata& \nodata &\small0.0070& \small0.0287& \small0.0900&\small16.6748\\
\small$f_{12}$&\nodata &\nodata &\nodata &\nodata &\small0.0057& \small0.0236& \small0.3752&\small13.7059\\
\small$f_{13}$&\nodata &\nodata &\nodata &\nodata & \small0.0065& \small0.0242& \small0.2452&\small14.0897\\
\small$f_{14}$&\nodata &\nodata &\nodata &\nodata & \small0.0006& \small0.0206& \small0.8211&\small11.9582\\

\enddata
\end{deluxetable}
\clearpage

\begin{deluxetable}{cccccccccccc}
\tabletypesize{\scriptsize} \tablecaption{Parameters of the primary
stars \label{tbl-3}} \tablewidth{0pt} \tablehead{
\colhead{\small{Star}} & \colhead{\small{Frequency }} &
\colhead{\small{$R_{1}/A$}} &
\colhead{\small{$\rho_{1}/\rho_{\sun}$}} & \colhead{\small{Q}} &
\colhead{\small{Mode}}  & \colhead{\small{$M_{bol}$}}&
\colhead{\small{$L/L_{\sun}$}} &
\colhead{\small{$R/R_{\sun}$}} & \colhead{\small{$M/M_{\sun}$}}\\
\colhead{} & \colhead{\small{(c/d)}} & \colhead{} & \colhead{} &
\colhead{} & \colhead{}  & \colhead{\small{(mag)}}& \colhead{}&
\colhead{}& \colhead{} } \startdata \small77.7795.29 & \small0.0100&
\small0.47& \small8.47E-8& \small0.029& \small1H
 & \small-3.75& \small2502 & \small153 & \small0.31  \\
& &&&&&&&\\
\small77.8031.42 & \small0.0112 & \small0.44 & \small3.66E-8 &
\small0.017 & \small2H  & \small-4.46 &\small4839 &
\small213 &\small0.36\\
\enddata
\end{deluxetable}

\clearpage

\end{document}